\documentstyle[12pt,epsf]{article}
\newcommand{\beq}{\begin{eqnarray}}
\newcommand{\eeq}{\end{eqnarray}}

\newcommand{\as}{ \alpha_s }
\newcommand{\aem}{ \alpha_{em} }

\newcommand{\mw}{ M_W }
\newcommand{\paa}{\pi_1^{\pm}}
\newcommand{\pbb}{\pi_8^{\pm}}
\newcommand{\pcc}{\tilde{\pi}^{\pm}}

\newcommand{\mpaa}{m_{\pi_1}}
\newcommand{\mpbb}{m_{\pi_8}}
\newcommand{\mpcc}{m_{\tilde{\pi}}}

\newcommand{\fpit}{F_{ \tilde{\pi}}}

\def\etap{\eta^{\prime}}
\def\etapp{\eta^{(')}}

\def\nceff{N_c^{{\rm eff}} }

\newcommand{\tab}[1]{Table \ref{#1}}
\newcommand{\fig}[1]{Fig.\ref{#1}}

\newcommand{\non}{\nonumber\\ }
\title{{\bf $\etap K$ puzzle of $B $ meson decays and new physics effects
in the TC2 model}}
\author{ Zhenjun Xiao$^{(1,2)}$
Wenjun Li$^{1}$, Libo Guo$^{(3)}$ and Gongru Lu$^{(1)}$\\
 {\small 1. Department of Physics, Henan Normal
University, Xinxiang, 453002 P.R. China.} \thanks{Mailing address}\\
 {\small 2. Department of Physics, Peking University,
Beijing, 100871 P.R. China.} \\
 {\small 3. Department of Physics,
Wuhan University, Wuhan, 430000 P.R. China.} \\  }
\date{\today}

\begin{document}
\maketitle
\begin{abstract}
Using the low energy effective Hamiltonian with the generalized factorization,
we calculate the new physics contributions to $B \to \pi^+ \pi^-, K \pi$ and
$ K \etap$ in the Topcolor-assisted-Technicolor(TC2) model, and compare the
results with the available data. By using $F_0^{{\rm B\pi}}(0)=0.20\pm 0.04 $
preferred by the CLEO data of $B \to \pi^+ \pi^-$ decay,  we find that the new
physics enhancements to $B \to K \etap $ decays
are significant in size, $\sim 50\%$ with respect to the standard model predictions,
insensitive to the variations of input parameters and hence provide a simple and
plausible new physics interpretation for the observed  unexpectedly large
$B\to K \etap $ decay rates.
\end{abstract}

%%%%%%%%%%%%%%%%%%%%%%%

\vspace{0.5cm} \noindent

\newpage
In two-body charmless hadronic $B$ meson decays, new physics beyond the standard model (SM)
may manifest itself through large enhancements to those penguin-dominated decay modes:
decays which are expected to be rare in the SM are found to have large branching ratios.
These potential deviations may be induced  by the virtual effects of new physics
through penguin and/or box diagrams \cite{slac504,fj99,atwood98,xiao20}.

In the framework of the SM, the two-body charmless hadronic decays $B \to h_1 h_2$ [ where
$h_1$ and $h_2$ are the light pseudo-scalar (P) and/or vector(V) mesons ]  have been studied
systematically by many authors \cite{bh1h2,du97,ali98,ali9804,chen99}.
On the experimental side, fourteen $B_{u,d} \to PP, PV$ decay channels have been observed by
CLEO, BaBar and BELLE Collaboration \cite{cleo99,cleo9912,cleo2000,babar2000,belle2000}:
\beq
B\to \pi^\pm \pi^\mp,\;  K \pi,\;  K \etap,\;  \rho^\pm \pi^\mp,\;
\rho^0\pi^\pm,\;  \omega\pi^\pm,\;  K^*\eta,\;  K^\pm \pi^\mp,\;  \phi K^\pm\;.
\eeq

By comparing the theoretical predictions with experimental measurements one finds the
following main points:
\begin{itemize}

\item
 The effective Hamiltonian with generalized factorization approach generally works
well to interpret the observed pattern of branching ratios. The penguin effects are
clearly observed\cite{cleosum}.

\item
There may exist a problem to accommodate the data of $\pi\pi$ and $K\pi$
simultaneously\cite{cheng08}.

\item
The $\etap K$ puzzle:  the $B \to K\etap$ decay rates are much larger
than what one ordinarily expected in the SM \cite{cleo2000,cheng08}.

\end{itemize}

Since 1997, the unexpectedly large branching ratio of $B \to K \etap $ has
stimulated much interests in literature \cite{cheng98,kagan97,hz98,du98,cheng00a}.
In order to accommodate the data, one may need
an additional contribution unique to the $\etap $ meson in the framework of the SM,
or new physics enhancements from new physics models beyond the SM to explain the
$B \to K \etap$ puzzle \cite{cleo2000}.
In a previous paper\cite{epj0007a}, we considered the second possibility and calculated
the new physics effects on the branching ratios and CP-violating asymmetries of
fifty seven $B \to PP, PV$ decay modes in the Topcolor-assited Technicolor
(TC2) model\cite{hill95}, and found that the new physics
enhancement to the penguin-dominated decay modes can be significant.
In another paper\cite{xiao10326}, we calculated the new physics contributions
to branching ratios of seventy six $B \to h_1 h_2$ decay modes in the general
two-Higgs-doublet models (models I, II and III).

In this
letter, we concentrate on the new physics contributions to seven observed B decay modes:
$B \to \pi^\pm \pi^\mp, K \pi$ and $B \to K \etap $ in the TC2 model.
Particular attention is devoted to the details of $B \to K \pi$ and $B \to K \etap$
decays when a smaller $F_0^{B\pi}(0)=0.20\pm 0.04$ instead of the ordinary
$F_0^{B\pi}(0)=0.33$ are used.

The effective Hamiltonian for the two-body charmless decays $B \to h_1 h_2$ are
now known at next-to-leading order (NLO) \cite{buchalla96a,ali9804,chen99}.
The standard theoretical frame to calculate the inclusive three-body decays
$b \to s \bar{q} q $\footnote{For $b \to d \bar{q} q$ decays, one simply make the
replacement $s \to d$.} is based on the effective Hamiltonian\cite{ali9804},
\begin{equation}
{\cal H}_{eff}(\Delta B=1) = \frac{G_F}{\sqrt{2}} \left \{
\sum_{j=1}^2 C_j \left ( V_{ub}V_{us}^* Q_j^u  + V_{cb}V_{cs}^*
Q_j^c \right ) - V_{tb}V_{ts}^* \left [ \sum_{j=3}^{10}  C_j Q_j +
C_{g} Q_{g} \right ] \right \}~, \label{heff2}
 \end{equation}
where the operator basis contains the current-current operators $Q_{1,2}$,
the QCD penguin operators $Q_{3-6}$, the electroweak penguin operators
$Q_{9-10}$ and the chromo-magnetic dipole operator $Q_g$, the explicit
expressions can be found easily for example in Ref.\cite{ali9804}.
Following Ref.\cite{ali9804}, we also neglect the effects
of the electromagnetic penguin operator $Q_{7\gamma}$, and do not
consider the effect of the weak annihilation and exchange diagrams.
The coefficients $C_{i}$ in Eq.(\ref{heff2}) are the
well-known Wilson coefficients. Within the SM and at scale $M_W$,
the Wilson coefficients $C_1(M_W), \cdots, C_{10}(M_W)$ at NLO level and
$C_{g}(M_W)$ at LO level have been given for example in \cite{buchalla96a}.

Following the same procedure as in the SM, it is straightforward to calculate the new
$\gamma$-, $Z^0$- and gluonic penguin diagrams induced by the exchanges of unit-charged
scalars, the technipion $\pi^{\pm}_1,\pi^{\pm}_8$ and top-pion $\tilde{\pi}^{\pm}$ appeared
in the TC2 model \footnote{For more details of TC2 model\cite{hill95} and the corresponding
constraint on its parameter space from the data, one can see
Refs.\cite{hill95,lane96,buchalla96b}
and our previous paper \cite{epj0007a}. For the sake of simplicity, we here do not present
the details about the calculations of new penguin diagrams in the TC2 model, but the reader
can find them in Ref.\cite{epj0007a,epj991}. }.

After including the new physics (NP) contributions induced by new penguin diagrams,
the Wilson coefficients $C_i(\mw)$ $i=1,\cdots, 10$ at NLO level and $C_g$ at
leading-order (LO) can be written as
\beq
C_1(\mw) &=& 1 - \frac{11}{6} \; \frac{\as(\mw)}{4\pi}
               - \frac{35}{18} \; \frac{\aem}{4\pi} \, , \label{eq:c1mw}\\
C_2(\mw) &=&     \frac{11}{2} \; \frac{\as(\mw)}{4\pi} \, , \\
C_3(\mw) &=& -\frac{\as(\mw)}{24\pi} \left [ E_0(x_t) +E_0^{NP} -\frac{2}{3}
\right ]\non
&& +\frac{\aem}{6\pi} \frac{1}{\sin^2\theta_W}
             \left[ 2 B_0(x_t) + C_0(x_t) + C_0^{NP} \right] \, ,\\
C_4(\mw) &=& \frac{\as(\mw)}{8\pi} \left [ E_0(x_t)+E_0^{NP}
-\frac{2}{3} \right ] \, ,\\
C_5(\mw) &=& -\frac{\as(\mw)}{24\pi}
\left [E_0(x_t)+E_0^{NP} -\frac{2}{3} \right ] \, , \\
C_6(\mw) &=& \frac{\as(\mw)}{8\pi}
\left [E_0(x_t)+E_0^{NP} -\frac{2}{3} \right ] \, ,\\
C_7(\mw) &=& \frac{\aem}{6\pi} \left [ 4 C_0(x_t) + 4 C_0^{NP}
    + D_0(x_t) +D_0^{NP} -\frac{4}{9}\right ]\, , \\
C_8(\mw) &=& 0 \, , \\
C_9(\mw) &=& \frac{\aem}{6\pi} \left\{ 4C_0(x_t) + 4 C_0^{NP}
    +D_0(x_t) +D_0^{NP} -\frac{4}{9}\right. \non
&& \left. +  \frac{1}{\sin^2\theta_W} \left [ 10 B_0(x_t)
- 4 C_0(x_t) + 4 C_0^{NP} \right ]  \right\} \, ,\\
C_{10}(\mw) &=& 0 \, , \label{eq:cimw}\\
C_{g}(\mw) &=& -\frac{1}{2}\left [ E'_0(x_t) + {E'}_0^{NP} \right ]
\, ,\label{eq:c8gmw}
\eeq
where $x_t=m_t^2/M_W^2$, the functions $B_0(x)$, $C_0(x)$, $D_0(x)$, $E_0(x)$
and $E'_0$ are the familiar Inami-Lim functions which describe
the contributions from the $W$-penguin and Box diagrams in the SM and can be found,
for example, in Refs.\cite{buchalla96a,xiao20}.
The functions $C_0^{NP}$, $D_0^{NP}$, $E_0^{NP}$ and ${E'}_0^{NP}$
describe the new physics contributions to Wilson coefficients in the TC2 model
as given in Ref.\cite{epj0007a},
\beq
C_0^{NP} &=&\frac{1}{\sqrt{2} G_F \mw^2 }\left [
\frac{\mpcc^2}{4\fpit^2} T_0(y_t)
+ \frac{\mpaa^2}{3 F_\pi^2 } T_0(z_t)
+ \frac{8 \mpbb^2}{3 F_\pi^2 } T_0(\xi_t) \right]~, \label{eq:c0tc2}\\
D_0^{NP} &=& \left \{ \frac{1}{4 \sqrt{2} G_F \fpit^2 } F_0(y_t)
 + \frac{1}{3 \sqrt{2} G_F F_\pi^2 } \left [  F_0(z_t) +
 8 F_0(\xi_t) \right ]  \right\}~, \label{eq:d0tc2}\\
E_0^{NP} &=& \left \{ \frac{1}{4\sqrt{2} G_F \fpit^2 } I_0(y_t)
 + \frac{1}{3 \sqrt{2} G_F F_\pi^2 } \left [  I_0(z_t) +
 8 I_0(\xi_t) + 9 N_0(\xi_t)\right ]  \right\}, \label{eq:e0tc2}\\
{E'}_0^{NP}&=& \left \{ \frac{1}{8 \sqrt{2} G_F \fpit^2 } K_0(y_t)
 + \frac{1}{6 \sqrt{2} G_F F_\pi^2 } \left [  K_0(z_t) +
 8 K_0(\xi_t) + 9 L_0(\xi_t)\right ]  \right\}, \label{eq:e0ptc2}
\eeq
where $y_t=m_t^{*2}/\mpcc^2$ with $m_t^*=(1-\epsilon) m_t$,
$z_t=( \epsilon m_t)^2/\mpaa^2$,$\xi_t=( \epsilon m_t)^2/\mpbb^2$, and
\beq
T_0(x)&=&-\frac{x^2}{8 (1-x)} -\frac{x^2}{8 (1-x)^2} \log[x]~,\label{eq:t0x}\\
F_0(x)&=& \frac{47-79 x + 38 x^2}{108 (1-x)^3}
     + \frac{3 -6 x^2+4 x^3}{18 (1-x)^4}\log[x]~, \label{eq:f0tc2}\\
I_0(x)&=&  \frac{ 7-29 x + 16 x^2}{36 (1-x)^3}
    -\frac{3 x^2-2 x^3}{6 (1-x)^4} \log[x]~, \label{eq:i0xtc2} \\
K_0(x)&=& -\frac{5-19 x + 20 x^2}{6 (1-x)^3} +
    \frac{x^2-2 x^3}{(1-x)^4}\log[x]~, \label{eq:k0xtc2}\\
L_0(x)&=& -\frac{4-5x - 5x^2}{6(1-x)^3}  -
    \frac{x-2x^2}{(1-x)^4}\log[x]~, \label{eq:l0xtc2}\\
N_0(x)&=&  \frac{11-7x + 2x^2}{36(1-x)^3} + \frac{1}{6(1-x)^4}
    \log[x]~. \label{eq:n0xtc2}
\eeq
The first term in Eq.(\ref{eq:c0tc2}) arises from the top-pion penguins, while
the second and third term correspond to the color-singlet and color-octect
technipion penguin respectively. For all four functions, the top-pion penguins
always dominate absolutely\cite{epj0007a}.

In numerical calculations, we use the following parameters of the TC2 model as
input parameter. Since the new physics contributions from technipions $\paa$ and $\pbb$
are much smaller than those from top-pion $\pcc$ within the reasonable parameter
space, we here fix  $\mpaa=100$GeV and $\mpbb=200$GeV for the sake of simplicity:
\beq
\mpaa=100GeV,\; \mpbb=200GeV, \; \fpit=50 GeV, \; F_{\pi}=120 GeV,\;
\epsilon=0.05, \label{eq:tc2fix}
\eeq
where $F_{\pi}$ and $\fpit$ are the decay constants for technipions and
top-pions, respectively. For $\mpcc$, we consider the range of
$\mpcc=200 \pm 100 $ GeV to check the mass dependence of branching ratios
of two-body charmless hadronic B meson decays studied. All other relevant
input parameters, such as the quark masses and form factors, $etc.,$
are given in the Appendix.

Since the heavy charged pseudo-scalars appeared in TC2 model have
been integrated out at the scale $M_W$, the QCD running of the
Wilson coefficients $C_i(M_W)$ down to the scale $\mu=O(m_b)$
after including the NP contributions will be the same as in the SM.
In the NDR scheme, by using the input parameters as given in
Appendix and Eq.(\ref{eq:tc2fix}) and setting $\mu=2.5$ GeV,  we find that:
\begin{eqnarray}
&& C_1 =  1.1245,\ \   C_2 = - 0.2662, \ \ C_3  =  0.0195,\ \   C_4 = -0.0441,\non
&& C_5 =  0.0111, \ \ C_6  =  -0.0535,\ \  C_7  =  0.0026, \ \  C_8  = 0.0018,\non
&& C_9 =  -0.0175,\ \ C_{10} = 0.0049,  C_g^{{\rm eff}}  =  0.3735 \label{eq:cgmb2}
\end{eqnarray}
where $C_g^{{\rm eff} } = C_{g}+ C_5$.

In this letter, the generalized factorization ansatz\footnote{For recent
discussions about the generalized factorization approach, one can see
Ref.\cite{cheng99a} and reference therein.} as being used in
Ref.\cite{chen99,epj0007a} will be employed. For the studied seven $B$ meson decay
modes, we use the decay amplitudes as given in Ref.\cite{ali9804} without
further discussion about details. We focus on estimating the new physics effects on
these seven measured decay modes.

In the NDR scheme and for $SU(3)_C$, the effective Wilson coefficients
can be written as \cite{chen99}
\beq
C_i^{{\rm eff} } &=& \left [ 1 + \frac{\alpha_s}{4\pi} \, \left( \hat{r}_V^T +
 \gamma_{V}^T \log \frac{m_b}{\mu}\right) \right ]_{ij} \, C_j
 +\frac{\alpha_s}{24\pi} \, A_i' \left (C_t + C_p + C_g \right)
+ \frac{\alpha_{ew}}{8\pi}\, B_i' C_e ~, \label{eq:wceff}
\eeq
where $A_i'=(0,0,-1,3,-1,3,0,0,0,0)^T$, $B_i'=(0,0,0,0,0,0,1,0,1,0)^T$, the
matrices  $\hat{r}_V$ and $\gamma_V$ contain the process-independent
contributions from the vertex diagrams, and  can be found, for example,
in Refs.\cite{chen99,cheng99a}.
The function $C_t$, $C_p$, and $C_g$ describe the contributions arising
from the penguin diagrams of the current-current
$Q_{1,2}$, the QCD operators $Q_3$-$Q_6$, and the tree-level diagram of the
magnetic dipole operator $Q_{8G}$, respectively. The explicit expressions of
the functions $C_t$, $C_p$, and $C_g$ can be found for example in
Refs.\cite{chen99,epj0007a}.
We here  also follow the procedure of Ref.\cite{ali98} to include the contribution
of magnetic gluon penguin.

In the generalized factorization ansatz, the effective Wilson coefficients $C_i^{{\rm eff}}$
will appear in the decay amplitudes in the combinations,
\beq
a_{2i-1}\equiv C_{2i-1}^{{\rm eff}} +\frac{{C}_{2i}^{{\rm eff}}}{\nceff}, \ \
a_{2i}\equiv C_{2i}^{{\rm eff}}     +\frac{{C}_{2i-1}^{{\rm eff}}}{\nceff}, \ \ \
(i=1,\ldots,5) \label{eq:ai}
\eeq
where the effective number of colors $\nceff$ is treated as a free parameter
varying in the range of $2 \leq \nceff \leq \infty$, in order to
model the non-factorizable contribution to the hadronic matrix elements.
Although  $\nceff$ can in principle
vary from channel to channel, but in the energetic two-body hadronic B
meson decays, it is expected to be process insensitive as supported by
the data \cite{chen99}. As argued in Ref.\cite{cheng98},
$\nceff(LL)$ induced by the $(V-A)(V-A)$ operators can be rather
different from $\nceff(LR)$ generated by  $(V-A)(V+A)$ operators.
In this paper, however, we will simply assume that $\nceff(LL)\equiv \nceff(LR)=
\nceff$ and consider the variation of $\nceff$ in the range
of $2 \leq \nceff \leq \infty$ since we here focus on the calculation
of new physics effects.

In the B rest frame, the branching ratios ${\cal B}(B \to PP)$ can
be written as
\beq
{\cal B}(B \to X Y )=  \frac{1}{\Gamma_{tot}} \frac{|p|}{8\pi M_B^2}
|M(B\to XY)|^2~,\label{eq:brbpp}
\eeq
where $\Gamma_{tot}(B_u^-)=3.982 \times 10^{-13}$ GeV
and $\Gamma_{tot}(B_d^0)=4.252 \times 10^{-13}$GeV obtained by using
$\tau(B_u^-)=1.653 ps$ and $\tau(B_d^0)=1.548 ps$ \cite{pdg2000}, $p_B$ is the
four-momentum of the B meson, $M_B=5.279$ GeV is the mass of $B_u$ or $B_d$ meson,
and
\beq
|p| =\frac{1}{2M_B}\sqrt{[M_B^2 -(M_X + M_Y)^2] [ M_B^2 -(M_X-M_Y)^2 ]}
\label{eq:pxy}
\eeq
is the magnitude of momentum of particle X and Y in the B rest frame.

For the seven studied $B$ meson decay modes, currently available measurements from
CLEO, BaBar and Belle Collaborations \cite{cleo9912,cleo2000,babar2000,belle2000}
are as follows:
\beq
{\cal B}(B \to \pi^+ \pi^-)&=& \left \{\begin{array}{ll}
( 4.3 ^{+1.6}_{-1.5} \pm 0.5 )\times 10^{-6} & {\rm [CLEO]}, \\
( 9.3 ^{+2.8\; +1.2 }_{-2.1\; -1.4} )\times 10^{-6} & {\rm [BaBar]},  \\
\end{array} \right. \label{eq:brexp01} \\
{\cal B}(B \to K^+ \pi^0)&=& \left \{\begin{array}{ll}
( 11.6 ^{+3.0\; +1.4}_{-2.7\; -1.3} )\times 10^{-6} & {\rm [CLEO]}, \\
( 18.8 ^{+5.5}_{-4.9} \pm 2.3 )\times 10^{-6} & {\rm [Belle]},  \\
\end{array} \right. \label{eq:brexp11} \\
{\cal B}(B \to K^+ \pi^-)&=& \left \{\begin{array}{ll}
( 17.2 ^{+2.5}_{-2.4} \pm 1.2)\times 10^{-6} & {\rm [CLEO]}, \\
( 12.5 ^{+3.0\; +1.3}_{-2.6\; -1.7} \pm 2.3 )\times 10^{-6} & {\rm [BaBar]},  \\
( 17.4 ^{+5.1}_{-4.6} \pm 3.4)\times 10^{-6} & {\rm [BELLE]}, \\
\end{array} \right. \label{eq:brexp12} \\
{\cal B}(B \to K^0 \pi^+)&=&
( 18.2 ^{+4.6}_{-4.0} \pm 1.6)\times 10^{-6}\ \   {\rm [CLEO]},
\label{eq:brexp13}\\
{\cal B}(B \to K^0 \pi^0)&=& \left \{\begin{array}{ll}
( 14.6 ^{+5.9\; +2.4}_{-5.1\; -3.3} )\times 10^{-6} & {\rm [CLEO]}, \\
( 21  ^{+9.3\; +2.5}_{-7.8\; -2.3} )\times 10^{-6} & {\rm [BELLE]},  \\
\end{array} \right. \label{eq:brexp14} \\
{\cal B}(B \to K^+ \etap )&=& \left \{\begin{array}{ll}
( 80 ^{+10}_{-9} \pm 7 )\times 10^{-6} & {\rm [CLEO]}, \\
( 62 \pm 18 \pm 8 )\times 10^{-6} & {\rm [BaBar]},  \\
\end{array} \right. \label{eq:brexp16} \\
{\cal B}(B \to K^0 \etap )&=&
( 89 ^{+18}_{-16} \pm 9 )\times 10^{-6} \ \  {\rm [CLEO]}.
\label{eq:brexp18}
\eeq
The measurements of CLEO, BaBar and BELLE Collaboration are consistent with each
other within errors.

In Table 1, we present the theoretical predictions of
the branching ratios for the seven $B \to PP$ decay modes in the framework
of the SM and TC2 model by using the form factors from Baner, Stech and Wirbel (BSW)
model\cite{bsw87} and Lattice QCD/QCD sum rule (LQQSR) model
 \cite{flynn97} form factors, as listed in the first  and second entries respectively.
Theoretical predictions are made by using the central values of
input parameters as given in Eq.(\ref{eq:tc2fix}) and the Appendix,
and assuming $m_{\tilde{\pi}}=200$GeV and $\nceff=2, 3,
\infty$ in the generalized factorization approach.
The branching ratios collected in the tables are the averages of
the branching ratios of $B$ and anti-$B$ decays. The ratio $\delta
{\cal  B}$  describes the new physics correction on the decay
ratio and is defined as
\begin{equation}
\delta {\cal  B} (B \to XY) = \frac{{\cal  B}(B \to XY)^{TC2}
-{\cal  B}(B \to XY)^{SM}}{{\cal  B}(B \to XY)^{SM}} \label{eq:dbr}
\end{equation}

\begin{table}[tb]
\begin{center}
\caption{Branching ratios (in units of $10^{-6}$) of seven studied $B $ decay modes
in the SM and TC2 model by using the BSW and LQQSR form factors, with $k^2=m_b^2/2$,
$A=0.81$, $\lambda=0.2205$, $\rho=0.12$, $\eta=0.34$, $\nceff=2,\; 3,\; \infty$
and $\mpcc=200$ GeV,  and by employing generalized factorization approach. }
\label{bpp1}
\vspace{0.2cm}
\begin{tabular} {l ccc ccc ccc } \hline \hline
 &  \multicolumn{3}{c}{SM }&
\multicolumn{3}{c}{TC2}& \multicolumn{3}{c}{$\delta {\cal  B} \; [\%]$}    \\
\cline{2-10}
Channel & $2$& $3$ & $\infty$ & $2$& $3$ & $\infty$&$2$&$3$& $\infty$ \\ \hline
$B^0 \to \pi^+ \pi^-$       & $9.03$ &$10.3$&$12.9$&$9.20$ &$10.4$&$13.1$&$1.9$&$1.8$&$1.6$ \\
                            & $10.7$ &$12.2$&$15.4$&$10.9$ &$12.2$&$15.6$&$1.9$&$1.8$&$1.6$ \\
$B^+ \to K^+ \pi^0$         & $12.1$ &$13.5$&$16.7$&$19.6$ &$21.8$&$26.5$&$63 $&$61 $&$59 $ \\
                            & $14.3$ &$16.0$&$19.8$&$23.3$ &$25.8$&$31.4$&$63 $&$61 $&$58 $ \\
$B^0 \to K^+ \pi^-$         & $17.7$ &$19.6$&$23.8$&$24.2$ &$26.7$&$32.0$&$37 $&$36 $&$35 $ \\
                            & $21.0$ &$23.3$&$28.3$&$28.8$ &$31.8$&$38.1$&$37 $&$36 $&$35 $ \\
$B^+ \to K^0 \pi^+$         & $20.0$ &$23.3$&$30.7$&$27.8$ &$32.8$&$44.1$&$39 $&$41 $&$44 $\\
                            & $23.8$ &$27.7$&$36.5$&$33.0$ &$39.0$&$52.4$&$39 $&$41 $&$44 $\\
$B^0 \to K^0 \pi^0$         & $7.22$ &$8.25$&$10.6$&$7.88$ &$9.28$&$12.5$&$9.3$&$13 $&$18 $ \\
                            & $8.61$ &$9.85$&$12.6$&$9.44$ &$11.1$&$15.0$&$9.6$&$13 $&$18 $\\
$ B^+ \to  K^+ \eta^\prime$ & $22.9$ &$28.8$&$42.9$&$34.3$ &$42.1$&$60.2$&$50 $&$46 $&$40 $\\
                            & $26.3$ &$33.1$&$49.3$&$39.3$ &$48.3$&$69.2$&$50 $&$46 $&$40 $\\
$B^0 \to K^0 \eta^\prime$   & $22.0$ &$28.3$&$43.1$&$33.0$ &$41.5$&$61.4$&$50 $&$47 $&$43 $ \\
                            & $25.3$ &$32.4$&$49.5$&$37.9$ &$47.6$&$70.5$&$50 $&$47 $&$43 $ \\
\hline
\end{tabular}\end{center}
\end{table}

By comparing the theoretical predictions with data, the following points can be understood:
\begin{itemize}

\item
For $B_d^0 \to \pi^+\pi^-$ decay, the SM prediction is clearly larger than the CLEO measurement,
but agree with BaBar measurement, while the BaBar measurement has a larger error than CLEO.
The new physics contribution to this tree-dominated decay mode is negligibly
small.

\item
For four $B \to K \pi$ decays, the SM predictions are agree with experimental
measurements. In TC2 models, the theoretical predictions are generally larger
than the data but still agree with the data with $2\sigma$
errors\cite{epj0007a} since both the theoretical and experimental errors are
still large now.

\item
For $B_u^+ \to K^+\etap$ and $B_d^0 \to K^0 \etap$ decay, the SM predictions are clearly
much smaller than the data (especially the CLEO measurement). But the new physics enhancement
can make the theoretical predictions in the TC2 model become agree with CLEO/BaBar  data within
one standard deviation.

\end{itemize}

The unexpectedly large $B \to K \etap$ decay rates were firstly observed in 1997
\cite{cleo98},
and confirmed recently with the full CLEO II and II.V samples \cite{cleo9912}.
The earlier SM predictions in the range of $(1-2)\times 10^{-5}$ are too
small compared with experiment. In the framework of the SM, the $B \to K \etap$ decays can
be enhanced through \cite{cheng08} (i) the small running mass $m_s$ at the
scale $m_b$ \footnote{However, a rather small $m_s$ is not consistent with recent lattice
calculations.}, (ii) the sizable $SU(3)$ breaking in the decay constant $f_0$
and $f_8$, (iii) larger form factor $F_0^{B\etap}(0)$ due to the smaller $\eta-\etap$
mixing angle $-15.4^0$ rather than $\approx -20^\circ$, (iv) contribution from
the $\etap$ charm content, and (v)  constructive interference in tree
amplitudes.  However, as pointed out in Ref.\cite{kagan97,ali98}, the above
mentioned enhancement is partially washed out by the anomaly effects in the
matrix element of pseudoscalar densities, an effect overlooked before. As a
consequence, the net enhancement is not very large:
${\cal B}( B^{\pm} \to K^{\pm} \etap) =(40-50)\times 10^{-6}$ as given in
Ref.\cite{cheng00a}.

\begin{figure}[t] %fig.1
\vspace{-60pt}
\begin{minipage}[t]{0.90\textwidth}
\centerline{\epsfxsize=\textwidth \epsffile{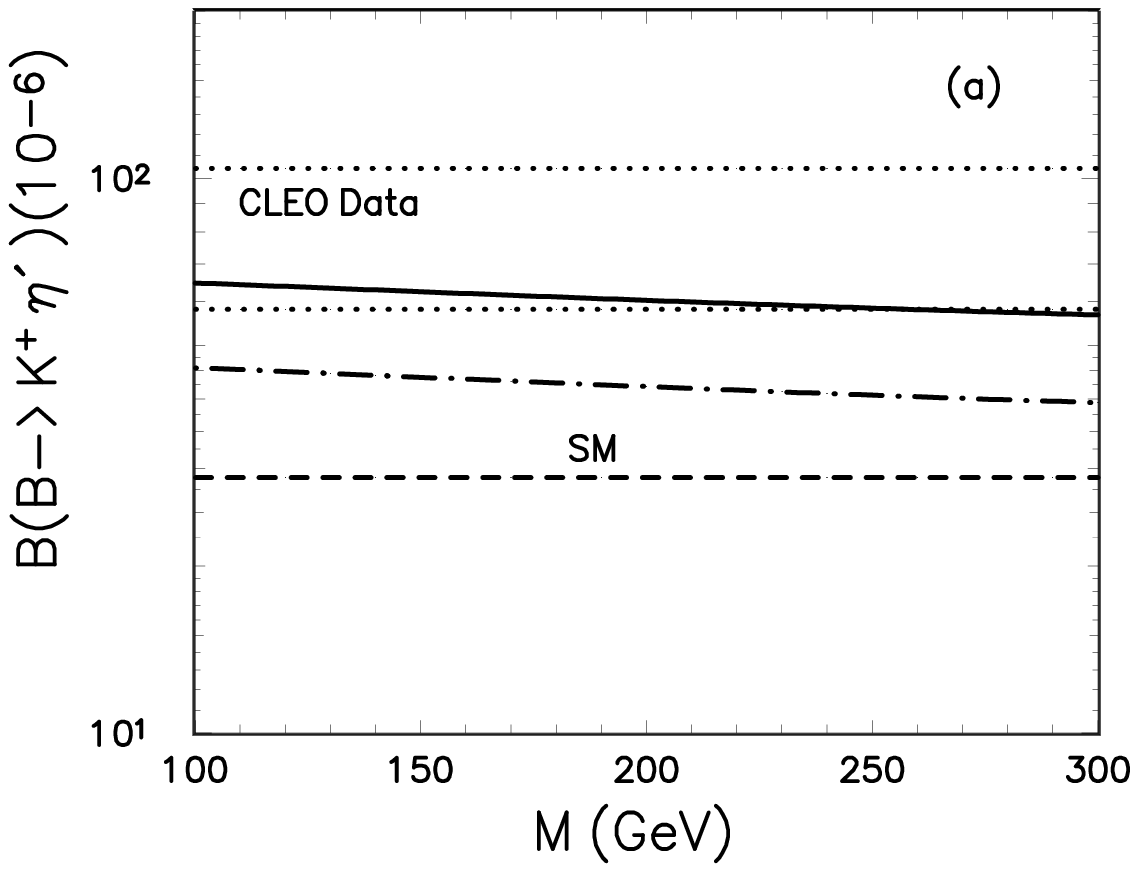}}
\vspace{-60pt}
\centerline{\epsfxsize=\textwidth \epsffile{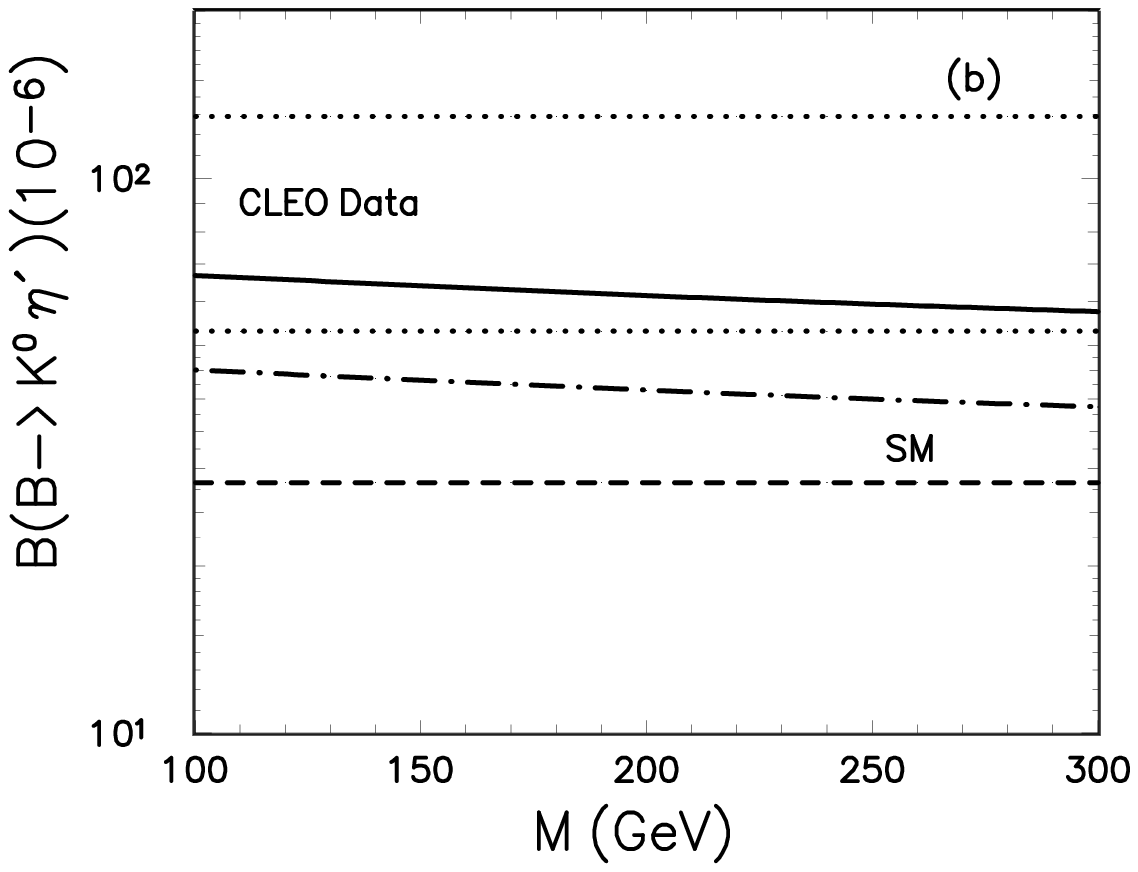}}
%\vspace{-20pt}
\caption{Plots of branching ratios of decays $B^+ \to K^+ \etap$ (1a) and
$B^0 \to K^0 \etap$ (1b) versus mass $\mpcc$ in the SM and TC2 model with
$F_0^{B\pi}(0)=0.33$ .
The short-dashed line shows the SM predictions with $\nceff=3$.
The dot-dashed and solid curve refers to the branching ratios in the TC2 model
for $\nceff=3$ and $\infty$, respectively. Theoretical uncertainties are not
shown here.
The dots band corresponds to the CLEO/BaBar data with $2\sigma$ errors:
${\cal  B}(B^\pm \to  K^\pm \etap)=(75  \pm 20 )\times 10^{-6}$ and
${\cal  B}(B^0 \to  K^0 \etap)=(89^{+40}_{-36})\times 10^{-6}$.}
\label{fig:fig1}
\end{minipage}
\end{figure}

In the TC2 model, on the other hand, the new gluonic and electroweak penguins contribute
through constructive interference with their SM counterparts and consequently provide
the large enhancements, $\sim 50\%$ with respect to the SM predictions, as shown
in \tab{bpp1}. By using $F_0^{B\pi}(0)=0.33$ and other input parameters as given in the
Appendix,  one  finds  numerically that
\beq
{\cal B}(B^{\pm} \to K^\pm \etap)&=& \left \{\begin{array}{ll}
( 20 - 52 )\times 10^{-6} & {\rm in \ \ SM}, \\
( 30 - 71 )\times 10^{-6} & {\rm in \ \ TC2},  \\
\end{array} \right. \label{eq:kpetap} \\
{\cal B}(B^0 \to K^0 \etap)&=& \left \{\begin{array}{ll}
( 19 - 52 )\times 10^{-6} & {\rm in \ \ SM}, \\
( 29 - 73 )\times 10^{-6} & {\rm in \ \ TC2},  \\
\end{array} \right. \label{eq:k0etap}
\eeq
where the effects induced by the uncertainties of major input parameters have been
taken into account.
The SM prediction is still smaller than the CLEO result but agree with the
BaBar measurement\footnote{One should note that the error of BaBar result is
much larger than that of CLEO result.}. In \fig{fig:fig1}, we plot the
mass-dependence of ${\cal B}(B^+ \to K^+ \etap)$ and ${\cal B}(B^0 \to K^0 \etap)$
in the SM and TC2 model. The short-dashed line in \fig{fig:fig1} shows the SM predictions
with $\nceff=3$. The dot-dashed and solid curve refers to the branching ratios in the
TC2 model for $\nceff=3$ and $\infty$, respectively.
The upper dots band corresponds to the data with $2\sigma$ errors:
${\cal  B}(B^\pm \to  K^\pm \etap)=(75  \pm 20 )\times 10^{-6}$ (average of CLEO and
BaBar result) and
${\cal  B}(B^0 \to  K^0 \etap)=(89^{+40}_{-36})\times 10^{-6}$ (CLEO only).
It is evident that the theoretical predictions in the TC2 model
are agree well with CLEO/BaBar data within one standard deviation.

Since $B_d^0 \to \pi^+\pi^-$ decay is a tree-dominated decay mode, the new physics
correction induced through loop diagrams should be very small, as shown in \tab{bpp1}.
The CLEO measurement of this mode puts a very stringent constraint on the form factor
$F_0^{B\pi}(0)$:  $F_0^{B\pi}(0)=0.20 \pm 0.04$ as given in
Ref.\cite{cheng9912}. In the SM, this smaller form factor will lead to two
difficulties:

\begin{enumerate}
\item
First, the predicted $B \to K \pi$ branching ratios
will be too small when compared with the data since their decay rates depend on
the  form factors $F_0^{B\pi}(0)$ and $F_0^{BK}(0)$.
We know that the form factor  $F_0^{BK}(0)$
cannot  deviate too much from $F_0^{B\pi}(0)$, otherwise the $SU(3)$-symmetry relation
$F_0^{B\pi}=F_0^{B\pi}$ will be badly broken.

\item
Second,  the predicted $B \to K \etap$ branching ratios will be also too small in
the SM since the branching ratio ${\cal B}(B \to K \etap)$ depends
on both the form factor $F_0^{B\pi}(0)$ and $F_0^{B\etap}$.
A small $F_0^{BK}(0)$ leads to a small $F_0^{B\etap}$ and in turn small branching
ratios of $B \to K \etap$ decays.
If we use the relation \cite{ali9804},
\beq
F_{0,1}^{B\etap}=F_{0,1}^{B\pi} \left ( \frac{\sin \theta_8}{\sqrt{6}}
+ \frac{\cos \theta_0}{\sqrt{3}} \right )\, ,\;
F_{0,1}^{B\eta}=F_{0,1}^{B\pi} \left ( \frac{\cos \theta_8}{\sqrt{6}}
- \frac{\sin \theta_0}{\sqrt{3}} \right )\,
\eeq
to define $F_0^{B\etap}$ with $\theta_0=-9,1^\circ$ and
$\theta_8=-22.2^\circ$\cite{ali9804}, the SM prediction for the branching ratio
${\cal B}(B \to K\etap)$ will be decreased by about $26\%$. In
\tab{bpp2}, we show the branching ratios of seven studied decay modes
obtained by using $F_0^{B\pi}(0)=0.20$ instead of $F_0^{B\pi}(0)=0.33$ while
keep all other input parameters remain the same as being used in \tab{bpp1}.

\end{enumerate}

\begin{figure}[t] %fig2
\vspace{-60pt}
\begin{minipage}[t]{0.90\textwidth}
\centerline{\epsfxsize=\textwidth \epsffile{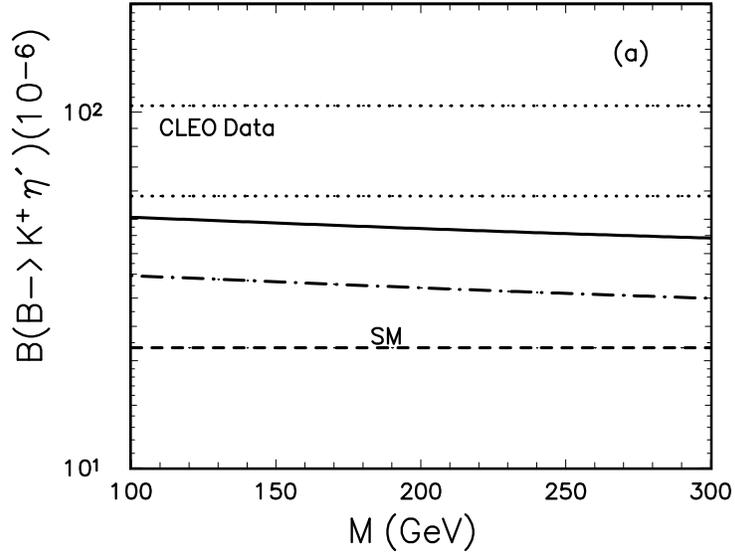}}
\vspace{-60pt}
\centerline{\epsfxsize=\textwidth \epsffile{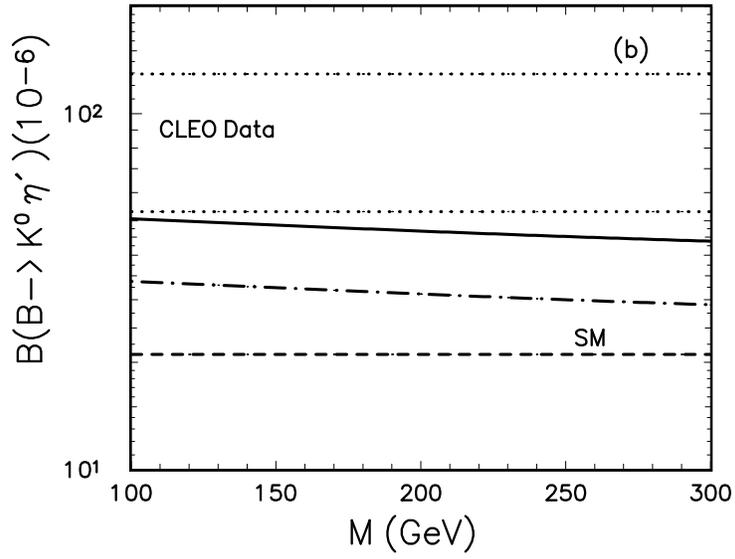}}
%\vspace{-20pt}
\caption{Same as \fig{fig:fig1} but for $F_0^{B\pi}(0)=0.20$ instead of $0.33$.}
\label{fig:fig2}
\end{minipage}
\end{figure}

In TC2 model, however, the decrease induced by using smaller $F_0^{B\pi}(0)$
will be compensated by large new physics enhancement and therefore restore
the agreement between the theoretical predictions and the data, as illustrated in
\fig{fig:fig2} for the decay $B \to K \etap$. By using $F_0^{B\pi}(0)=0.20\pm 0.04$,
one finds that
\beq
{\cal B}(B^{\pm} \to K^\pm \etap)&=& \left \{\begin{array}{ll}
( 12 - 45 )\times 10^{-6} & {\rm in \ \ SM}, \\
( 20 - 62 )\times 10^{-6} & {\rm in \ \ TC2},  \\
\end{array} \right. \label{eq:kpetapb} \\
{\cal B}(B^0 \to K^0 \etap)&=& \left \{\begin{array}{ll}
( 11 - 44 )\times 10^{-6} & {\rm in \ \ SM}, \\
( 20 - 61 )\times 10^{-6} & {\rm in \ \ TC2},  \\
\end{array} \right. \label{eq:k0etapb}
\eeq
where the effects of major uncertainties have been taken into account.

In \fig{fig:fig2}, we plot the mass dependence of ${\cal B}(B^+ \to K^+
\etap)$ and ${\cal B}(B^0 \to K^0 \etap)$ in the SM and TC2 model by using
$F_0^{B\pi}(0)=0.20$ instead of $0.33$ (while all other input parameters are the same as in
\fig{fig:fig1}). The short-dashed line in \fig{fig:fig2} shows the SM predictions
with $\nceff=3$. The dot-dashed and solid curve refers to the branching ratios in the
TC2 model for $\nceff=3$ and $\infty$, respectively.
The upper dots band corresponds to the CLEO/BaBar data with $2\sigma$ errors.
It is easy to see that (a) the gap between the SM predictions of $B \to K \etap$ decay
rates and the data is enlarged by using $F_0^{B\pi}(0)=0.20$ instead of $0.33$, and
(b) the new physics enhancement therefore becomes essential for the theoretical
predictions to be consistent with CLEO/BaBar result within $2\sigma$ errors.

\begin{table}[tb]
\begin{center}
\caption{Branching ratios (in units of $10^{-6}$) of seven studied $B $ decay modes
in the SM and TC2 model by using the BSW form factors with $F_0^{B\pi}(0)=0.20$ instead of
$F_0^{B\pi}(0)=0.33$, assuming $k^2=m_b^2/2$,
$A=0.81$, $\lambda=0.2205$, $\rho=0.12$, $\eta=0.34$, $\nceff=2,\; 3,\; \infty$
and $\mpcc=200$ GeV,  and by employing generalized factorization approach.}
\label{bpp2}
\vspace{0.2cm}
\begin{tabular} {l ccc ccc ccc } \hline \hline
 &  \multicolumn{3}{c}{SM }&
\multicolumn{3}{c}{TC2}& \multicolumn{3}{c}{$\delta {\cal  B} \; [\%]$}    \\
\cline{2-10}
Channel & $2$& $3$ & $\infty$ & $2$& $3$ & $\infty$&$2$&$3$& $\infty$ \\ \hline
$B^0 \to \pi^+ \pi^-$       & $3.32$ &$3.77$&$4.75$&$3.38$ &$3.83$&$4.83$&$ 1.9$&$ 1.8$&$ 1.6$ \\
$B^+ \to K^+ \pi^0$         & $5.12$ &$5.77$&$7.25$&$9.09$ &$10.1$&$12.5$&$ 77.6$&$ 75.8$&$ 71.9$ \\
$B^0 \to K^+ \pi^-$         & $6.49$ &$7.20$&$8.73$&$8.90$ &$9.80$&$11.7$&$ 37.1$&$ 36.2$&$ 34.6$ \\
$B^+ \to K^0 \pi^+$         & $7.34$ &$8.55$&$11.3$&$10.2$ &$12.0$&$16.2$&$ 38.9$&$ 40.7$&$ 43.7$\\
$B^0 \to K^0 \pi^0$         & $2.20$ &$2.47$&$3.11$&$1.94$ &$2.27$&$3.07$&$-12.0$&$-8.2$&$-1.5$ \\
$ B^+ \to  K^+ \eta^\prime$ & $16.9$ &$21.7$&$33.3$&$25.6$ &$32.0$&$47.0$&$ 51.7$&$ 47.3$&$ 41.0$\\
$B^0 \to K^0 \eta^\prime$   & $16.2$ &$21.0$&$32.8$&$24.5$ &$31.1$&$46.7$&$ 51.7$&$ 47.8$&$ 42.5$ \\
\hline
\end{tabular}\end{center} \end{table}

From \tab{bpp2}, it is easy to see that the new physics enhancement to first
three $B \to K \pi$ decays and $B \to K \etap$ decays are still large in size and play
an important role to boost the corresponding branching ratios to be consistent
with experimental measurements.

For the $B \to K^0 \pi^0$ decay mode, the SM predictions are always smaller
than the data although the error of the data is still very large, as can be seen from
Tables 1-2. For the case of using $F_0^{B\pi}(0)=0.33$, the new physics contribution
in TC2 model provide a $(10-20)\%$ enhancement. For the case of using
$F_0^{B\pi}(0)=0.20$, however, the new physics contribution in TC2 model result in
a $(2-12)\%$ decrease. We currently are not sure that whether there is a
discrepancy between the theory and the data for this decay mode. This is an open
problem now, further refinement of the data will clear this point soon.

In short, we here studied the new physics contributions to the seven observed $B\to PP$
decay modes by employing the effective Hamiltonian with generalized factorization.
In this letter, particular attention is devoted to the details of $B \to K \etap$
decays, and to the discussions about currently known mechanisms to enhance this
decay mode. We made the numerical calculation by using both $F_0^{B\pi}(0)=0.33$ as
given in the ordinary BSW model, as well as $F_0^{B\pi}(0)=0.20\pm 0.04 $ preferred by
the CLEO data of $B \to \pi^+ \pi^-$. We presented the numerical results in Tables 1-2
and Figs.1-2. We also discussed the difficulties induced by using the
smaller $F_0^{B\pi}(0)$ and shown that one can accommodate the data of
$\pi^+\pi^-$, $K^+\pi$, and $K^0 \pi^+$ simultaneously after taking into account
the new physics contributions. But we are still not sure if there is a
discrepancy between the theory and the data for $B \to K^0 \pi^0$ decay mode.

By using whether $F_0^{B\pi}(0)=0.20\pm 0.04$ or $0.33$, we always found that the
new physics enhancements to $B \to K \etap$ decays are significant in size, and
hence the theoretical predictions of ${\cal B}(B \to K \etap)$ in the TC2 model
are agree with CLEO/BaBar data within $2\sigma$ errors. This seems to be a simple
and plausible new physics interpretation for the observed $\etap K$ puzzle.

\section*{Appendix:  Input parameters} \label{app:a}

In this appendix we present the relevant input parameters.

\begin{itemize}

\item
Input parameters of electroweak and strong coupling constant,
gauge boson masses, light meson masses, $\cdots$,
are as follows (all masses in unit of GeV )\cite{ali9804,pdg2000}
\beq
&& \alpha_{em}=1/128, \;  \alpha_s(M_Z)=0.118,\; \sin^2\theta_W=0.23,\; G_F=1.16639\times 10^{-5} (GeV)^{-2}, \non
&& M_Z=91.188, \;   M_W=80.42,\; m_{B_d^0}=m_{B_u^\pm}=5.279,\; m_{\pi^\pm}=0.140,\;\non
&& m_{\pi^0}=0.135,\;   m_{\eta}=0.547,\; m_{\etap}=0.958,\; m_{K^\pm}=0.494,\;  m_{K^0}=0.498.
\label{masses}
\eeq

\item
For the elements of CKM matrix, we use Wolfenstein parametrization, fix the
parameters $A, \lambda$ and $\rho$ to their central values:
$ A=0.81,\; \lambda=0.2205, \; \rho=0.12$, but varying $\eta$ in the range of
$\eta=0.34 \pm 0.08$.

\item
We first treat the internal quark masses in the loops as constituent masses:
$m_b=4.88 {\rm GeV}, \; m_c=1.5 {\rm GeV},\; m_s=0.5 {\rm GeV}, \; m_u=m_d=0.2 {\rm GeV}$.
Second, we use the current quark masses for $m_i$ ($i=u,d,s,c,b$) which appear through
the equation of motion when working out the hadronic matrix elements. For $\mu=2.5
{\rm GeV}$, one found \cite{ali9804}: $ m_b=4.88 {\rm GeV}, m_c=1.5 {\rm GeV},
m_s=0.122 {\rm GeV}, \; m_d=7.6 {\rm MeV},\; m_u=4.2 {\rm MeV}$.
For the mass of heavy top quark we  use $m_t=\overline{m_t}(m_t)=168 {\rm GeV}$.

\item
For the decay constants of light mesons, the following values will
be used in the numerical calculations (in the units of MeV):
\beq
&&f_{\pi}=133,\; f_{K}=158,\; f^u_{\eta}=f^d_{\eta}=78,\;
f^u_{\etap}=f^d_{\etap}=68,\non
&&f^c_{\eta}=-0.9,\; f^c_{\etap}=-0.23, f^s_{\eta}=-113,\; f^c_{\etap}=141. \label{fpis}
\eeq
where $f^u_{\etapp}$ and $f^s_{\etapp}$ have been defined in the two-angle-mixing
formalism with $\theta_0=-9.1^\circ$ and $\theta_8 =-22.2^\circ$\cite{fk97}.

\item
The relevant form factors are \cite{ali9804}
\beq
F_0^{B\pi}(0)=0.33,\; F_0^{BK}(0)=0.38,\; F_0^{B\eta}(0)=0.145,\;
F_0^{B\etap}(0)=0.135,
\eeq
in the BSW model \cite{bsw87}, and
\beq
F_0^{B\pi}(0)=0.36,\; F_0^{BK}(0)=0.41,\;
F_0^{B\eta}(0)=0.16,\; F_0^{B\etap}(0)=0.145,
\eeq
in the LQQSR approach.
And the momentum dependence of form factor $F_0(k^2)$ was defined in Ref.\cite{bsw87} as
$ F_0(k^2)= F_0(0)/(1-k^2/m^2(0^+))$. The pole masses being used to evaluate the
$k^2$ dependence of form factors are $m(0^+)=5.73$ GeV for $\bar{u}b$ and $ \bar{d}b$
currents, and $ m(0^+)= 5.89$ GeV for $\bar{s}b $ currents.

\end{itemize}

\section*{ACKNOWLEDGMENTS}

Authors are very grateful to D.S. Du, K.T. Chao, C.S.Li, C.D. L\"u,
Y.D. Yang and M.Z.Yang for helpful discussions.
Z.J. Xiao acknowledges the support by the National
Natural Science Foundation of China under the Grant No.19575015 and 10075013,
the Excellent Young Teachers Program of Ministry of Education, P.R.China and the
Natural Science Foundation of Henan Province under the Grant No. 994050500.

\newpage

\newpage

\end{document}